\documentclass[12pt]{iopart}
\usepackage{iopams}

\newcommand{\be}{\begin{equation}}
\newcommand{\ee}{\end{equation}}
\newcommand{\bea}{\begin{eqnarray}}
\newcommand{\eea}{\end{eqnarray}}
\newcommand{\bml}{\numparts}
\newcommand{\eml}{\endnumparts}

\newcommand{\vx}{\vec{x}}

\newcommand{\vz}{\vec{z}}
\newcommand{\vk}{\vec{k}}

\newcommand{\vq}{\vec{q}}
\newcommand{\vp}{\vec{p}}

\newcommand{\ep}{\epsilon}

\newcommand{\Ket}[1]{\left|\, #1 \,\right\rangle }
\newcommand{\Bra}[1]{\left\langle #1 \right|}
\newcommand{\Bracket}[2]{\langle\, #1 \,|\, #2\,\rangle }
\newcommand{\ma}[1]{\mathrm{#1}}
\newcommand{\inte}[1]{\int \rmd^{3}#1}

\begin{document}

\letter{The noncommutative degenerate electron gas}

\author{F S Bemfica\dag\ and H O Girotti\dag}

\address{\dag\ Instituto de F\'{\i}sica, Universidade Federal do Rio Grande do Sul,
Caixa Postal 15051, 91501-970 - Porto Alegre, RS, Brazil}

\eads{\mailto{fbemfica@if.ufrgs.br}, \mailto{hgirotti@if.ufrgs.br}}

\begin{abstract}
The quantum dynamics of nonrelativistic single particle systems involving
noncommutative coordinates, usually referred to as noncommutative quantum mechanics,
has lately been the object of several investigations. In this note we pursue these
studies for the case of multi-particle systems. We use as a prototype the degenerate
electron gas whose dynamics is  well known in the commutative limit. Our central aim here is to understand qualitatively, rather than quantitatively, the main modifications induced by the presence of noncommutative coordinates. We shall first
see that the noncommutativity modifies the exchange correlation energy while
preserving the electric neutrality of the model. By employing time-independent
perturbation theory together with the Seiberg-Witten map we show, afterwards, that
the ionization potential is modified by the noncommutativity. It also turns out that
the noncommutative parameter acts as a reference temperature. Hence, the
noncommutativity lifts the degeneracy of the zero temperature electron gas.

\end{abstract}


\submitto{\JPA}

\maketitle


The first paper on quantum field theories formulated in a noncommutative space-time
manifold was published in 1947 \cite{Snyder1}, although the idea that a
noncommutative space-time manifold might provide a solution for the problem of
ultraviolet divergences seems to have been suggested long before \cite{Jackiw1}. The
subject was, perhaps, abandoned due to the success of renormalization theory and its
revival is rather recent and related to string theory. Indeed, the noncommutative
Yang-Mills theory arises as a limit of string theory \cite{Connes1} and it was
extracted by Seiberg and Witten \cite{Seiberg1} by starting from the open string in
the presence of a magnetic field. More details on this and related subjects can be
found in the already existing review articles
\cite{Rivelles1,Douglas1,Szabo1,Gomes1,Girotti1,Girotti2} and also in the
specialized literature.

On the other hand, noncommutative quantum mechanics has also been under scrutiny
\cite{Chaichian1,Gamboa1,Gamboa2}. The main outcome, in the case of single particle
systems, is that a modification of the equal-time algebra obeyed by the basic
position observables acts as a source of new interactions which may or may not
preserve the original symmetries. The present paper is dedicated to study the physical consequences of introducing noncommutative coordinates in the case of quantum many-particle systems.

We consider, as a prototype, an idealized high density degenerate electron gas
occupying a volume $V = L^3$. As it is currently assumed, the electrons (electric
charge $-e$) are in the presence of a uniform background of positive ions (electric
charge $+e$) that makes the whole system electrically neutral. For this to be the
case the number of electrons must equal the number of ions ($N$). The ions are much
heavier than the electrons and will, then, be considered as static. Although this
system has been extensively described in textbooks \cite{Fetter} we make a small
digression here to pinpoint its highlights. By election, the degrees of freedom of
each electron ($E$) are the Cartesian positions $\{X^j_{a\,,\,E}\}$ and linear
momenta $\{P^j_{a\,,\,E}\}$, together with the spins $\{S^j_{a\,,\,E}\}$. For the
ions ($B$), the corresponding observables are, respectively, $\{X^j_{a\,,\,B}\}$,
$\{P^j_{a\,,\,B}\}$ and $\{S^j_{a\,,\,B}\}$. Small case letters from the beginning
of the Latin alphabet ($a, b, \ldots$) designate the particle, while small case
letters from the middle of the Latin alphabet ($i, j, \ldots$) only run from $1$ to
$3$ and identify the Cartesian component of the corresponding observable.
Observables associated with the electrons commute with those associated with the
ions. The phase space electron degrees of freedom obey the standard equal-time
algebra

\bea\label{1}
\eqalign{&\left[X^i_{a\,,\,E}\,,\,X^j_{b\,,\,E}\right]\,=\,0\,,\\
&\left[X^i_{a\,,\,E}\,,\,P^j_{b\,,\,E}\right]\,=\,\rmi\,\hbar\,\delta^{ij}\,\delta_{ab}\,,\\
&\left[P^i_{a\,,\,E}\,,\,P^j_{b\,,\,E}\right]\,=\,0\,.}
\eea

\noindent
We emphasize that all position observables commute among themselves. The equal-time
algebra for the ion phase space variables can be obtained from Eq.(\ref{1}) just by
replacing $E$ by $B$. The algebra of the spin components will not be explicitly
displayed.

Structurally, the more general form of the total Hamiltonian reads

\be
\label{2}
H\,=\,H_{E}\,+\,H_{B}\,+\,H_{EB}\,.
\ee

\noindent
The Hamiltonian $H_{E}$ describes the free dynamics of the electrons plus the
Coulomb interaction among them. Hence, in the position representation
($X^i\,\Ket{\vx}\,=\,x^i\,\Ket{\vx}\,,\,P^i \rightarrow p^i\,=\,-\,\rmi\,\hbar\,
\partial / \partial x^i$),

\bea
\label{3}
H_{E}\,=\,\sum_{a=1}^{N}\frac{ p^i_{a,E}\,p^i_{a,E}}{2m}\,+\,\frac{1}{2}\,\sum_{a\ne b}^{N}\, \ma{V}(|\, {\vec x}_{a\,,\,E}\,-\,{\vec x}_{b\,,\,E}\,|)\,,
\eea

\noindent
where

\be
\label{4}
\ma{V}(| {\vec r}|)\,=\,e^2\,\,\frac{\rme^{-\mu\,| {\vec r}|}}{| {\vec r}|}
\ee

\noindent
and $\mu$ is a damping factor needed to secure, in the thermodynamic limit, the
convergence of each term in the right hand side of Eq.(\ref{2}) \cite{Fetter}. As
for $H_{B}$ one writes

\bea
\label{5}
H_{B}\,&=&\,\frac{1}{2}\,\sum_{a\ne b}^{N}\,\ma{V}(|\, {\vec x}_{a\,,\,B}\,-\,{\vec x}_{b\,,\,B}\,|)\nonumber\\
&\rightarrow&\,  \frac{1}{2}\,\int \rmd^3x_{a\,,\,B}\,\int \rmd^3x_{b\,,\,B}\,n({\vx}_{a\,,\,B})\,n({\vx}_{b\,,\,B})\,\ma{V}(|\, {\vec x}_{a\,,\,B}\,-\,{\vec x}_{b\,,\,B}\,|).
\eea

\noindent
The absence of kinetic term in Eq.(\ref{5}) reflects the fact that the ions are
static. Furthermore, the continuous nature of the ion background is taken into
account by replacing the discrete summations by continuous integrals. This brings
into play the new variable, $n({\vx}_{a\,,\,B})$, known as the ion density. Finally,
$H_{EB}$ is taken to be

\bea
\label{6}
H_{EB}\,&=&\,-\sum_{a=1}^{N}\sum_{b=1}^N\,\ma{V}(|\, {\vec x}_{a\,,\,E}\,-\,{\vec x}_{b\,,\,B}\,|)\nonumber\\
&\rightarrow& \,-\sum_{a=1}^{N}\int \rmd^3x_{b\,,\,B}\,n({\vx}_{b\,,\,B})\,\ma{V}(|\, {\vec x}_{a\,,\,B}\,-\,{\vec x}_{b\,,\,B}\,|).
\eea

\noindent
We shall always be working in the approximation $n({\vx}_{a\,,\,B}) = constant = N/V$.

It has long been shown \cite{Fetter} that $H$, in Eq.(\ref{2}), can be cast

\be
\label{7}
H\,=\,H_0\,+\,H_I\,,
\ee

\noindent
where

\bml
\label{8}
\bea
H_0\,&=&\,\sum_{\vec{k}\lambda}\frac{\hbar^2k^2}{2m}\,c^\dagger_{\vec{k}\lambda}\,c_{\vec{k}\lambda}\,,\label{mlett:a8}\\
H_I\,&=&\,\frac{2\pi}{V}\,\sum_{\vec{k}\vec{p}\vec{q}}{}^{\prime}
\sum_{\lambda_1\lambda_2}\frac{e^2}{q^2}\,c^{\dagger}_{\vec{k}+\vec{q}\,,\,\lambda_1}\,c^{\dagger}_{\vec{p}-\vec{q}\,,\,\lambda_2}\,c_{\vec{p}\,,\,\lambda_2}\,c_{\vec{k}\,,\,\lambda_1}\,.\label{mlett:b8}
\eea
\eml

\noindent
Here, $c^\dagger_{\vec{k}\lambda}$ ($c_{\vec{k}\lambda}$) are the creation
(annihilation) operators of electrons of momentum $\vk$ and spin $\lambda$, whereas
$q \equiv |\vq|$. Furthermore, the prime in the summation symbol implies that the
momentum ${\vec q} = 0$ is excluded. Within the framework of time-independent
perturbation theory, the main outcomes, including contributions up to the second
order, may be summarized as follows. The unperturbed ground state (Fermi) energy
($E_0^{(0)}$) is given by \cite{Fetter}

\be
\label{9}
E_0^{(0)}\,=\,\frac{e^2}{2 a_0}\,N\,\frac{2.21}{r_s^2}\,,
\ee

\noindent
where $a_0$ is the Bohr radius, $r_s \equiv r_0/a_0$ and $\frac{4}{3}\pi r_{0}^{3} =
V/N$. Moreover, its first and second order perturbative corrections were,
respectively, found to read \cite{Fetter,Wigner,Gellmann1,Gellmann2}

\be
\label{10}
E_0^{(1)}\,=\,-\,\frac{e^2}{2 a_0}\,N\,\frac{0.916}{r_s}\,,
\ee

\noindent
and

\be
\label{11}
E_0^{(2)}\,=\,\frac{e^2}{2
a_0}\,N\,\left[\ep_0^{(2)r}\,+\,\ep_0^{(2)b}\right]\,=\,\frac{e^2}{2
a_0}\,N\,\left[0.0622\,\ln r_s\,-\,0.094\right]\,.
\ee

\noindent
Here \cite{Onsager},

\bea\label{12}
\fl\ep_0^{(2)b}\,=\,\frac{3}{16\pi^{5}}\int\frac{\rmd^{3}{\bf q}}{{\bf q}^{2}}\,\int_{|\vec{\bf k}+\vec{\bf q}|>1}\rmd^{3}{\bf k}\,\int_{|\vec{\bf p}-\vec{\bf q}|>1}\rmd^{3}{\bf p}\,\frac{\xi(1-{\bf k})\,\xi(1-{\bf p})}{[{\vec{\bf  q}\cdot}({\vec{\bf q}}+\vec{\bf k}-\vec{\bf p})](\vec{\bf q}+\vec{\bf k}-\vec{\bf p})^{2}}\nonumber\\
\lo=\,\frac{1}{3}\,\ln 2\,-\,\frac{3}{2 \pi^2}\,\zeta(3)\,\approx\,0.048\,,
\eea

\noindent
is the exchange correlation energy in Rydberg's units. We shall designate by
$\xi(x)$ the Heaviside step function. In order to work with dimensionless vectors
($\vec {\bf p}$) we define $k_F\,\vec {\bf p} \equiv \vp$, where $k_{F} \equiv
\left(\frac{9 \pi}{4}\right)^{ \frac{1}{3}}\,r_0^{- 1}$ is the Fermi wavenumber.
Also, ${\bf p} \equiv |\vec {\bf p}|$. Here ends our brief summary about the
degenerate electron gas.

We turn next into studying the implications of replacing $X^i_{a\,,\,E} \rightarrow
Q^i_{a\,,\,E},\,P^i_{a\,,\,E} \rightarrow P^i_{a\,,\,E}$, now obeying the equal-time
phase space algebra

\bea\label{13}
\eqalign{&\left[Q^i_{a\,,\,E}\,,\,Q^j_{b\,,\,E}\right]\,=\,2\,\rmi\,\delta_{a\,,\,b}\,\Theta^{ij}_E\,,\\
&\left[Q^i_{a\,,\,E}\,,\,P^j_{b\,,\,E}\right]\,=\,\rmi\,\hbar\,\delta^{ij}\,\delta_{ab}\,,\\
&\left[P^i_{a\,,\,E}\,,\,P^j_{b\,,\,E}\right]\,=\,0\,.}
\eea

\noindent
The distinctive feature of the new position observables ($Q^i_{a\,,\,E}$) is that
they do not commute among themselves. This lack of noncommutativity is characterized
by the real antisymmetric constant matrix ($\Theta^{ij}_E$). An explicit
representation for this algebra has already been obtained
\cite{Chaichian1,Gamboa1,Gamboa2,Girotti2} after realizing that (see Eqs.(\ref{1})
and (\ref{14}))

\be
\label{14}
Q^i_{a\,,\,E}\,=\,X^i_{a\,,\,E}\,-\,\frac{1}{\hbar}\,\Theta^{ij}_E\,P^j_{a\,,\,E}\,.
\ee

\noindent
A similar modification should be introduced for the ions. However, for static ions
Eq.(\ref{14}) reduces to $Q^i_{a\,,\,B}\,=\,X^i_{a\,,\,B}$.

As for the Hamiltonians, the replacement $X^i_{a\,,\,E} \rightarrow
Q^i_{a\,,\,E},\,P^i_{a\,,\,E} \rightarrow P^i_{a\,,\,E}$ amounts to $H \rightarrow
{\cal H}$, such that

\be
\label{15}
{\cal H}\,=\,{\cal H}_{E}\,+\,{\cal H}_{B}\,+\,{\cal H}_{EB}\,,
\ee

\noindent
where

\bea
\label{16}
{\cal H}_{E}\,=\,\sum_{a=1}^{N}\frac{p^i_{a,E}\,p^i_{a,E}}{2m}\,+\,
\frac{1}{2}\,\sum_{a\ne b}^{N}\, \ma{V}\left(\left|\,\,\vec{{\phi}}_{a\,,\,E}\,-\,\vec{{\phi}}_{b\,,\,E}\,\,\right|\right)\,,
\eea

\noindent

\bea
\label{17}
{\cal H}_{B}\,&=&\, \frac{1}{2}\,\int \rmd^3x_{a\,,\,B}\,\int \rmd^3x_{b\,,\,B}\,n({\vx}_{a\,,\,B})\,n({\vx}_{b\,,\,B})\,\ma{V}(|\, {\vec x}_{a\,,\,B}\,-\,{\vec x}_{b\,,\,B}\,|)\nonumber\\
&=&\,H_B\,,
\eea

\noindent
and

\be
\label{18}
{\cal H}_{EB}\,=\,-\sum_{a=1}^{N}\int \rmd^3x_{b\,,\,B}\,n({\vx}_{b\,,\,B})\,
\ma{V}\left(\left|\,\,\vec{\phi}_{a\,,\,E}\,-\,\vx_{b\,,\,B}\,\,\right|\right).
\ee

\noindent
For simplifying purposes, we have introduced the notation

\be
\label{19}
\phi^i_{a\,,\,E}\,\equiv\,x^i_{a\,,\,E}\,-\,\frac{1}{\hbar}\,\Theta^{ij}_E\,p^i_{a\,,\,E}\,.
\ee

\noindent
As already stated, $\{x^i_{a\,,\,E}\}$ denotes the set of eigenvalues of the
operator $X^i_{a\,,\,E}$, whereas  $p^i_{a\,,\,E}\,\equiv\,- \rmi \hbar
\partial/\partial x^i_{a\,,\,E}$ represents $P^i_{a\,,\,E}$ in the basis defined by
the common eigenvectors of $\{X^i_{a\,,\,E}\}$.

We next focus on $\ma{V}\left(\left|\vec{\phi}_{a\,,\,E}\,-\,\vx_{b\,,\,B}\right|\right)$
when acting on an arbitrary but differentiable function
$\Psi(x^i_{a\,,\,E}\,,\,x^i_{b\,,\,B})$. By taking into account Eq.(\ref{19}) one
finds that

\bea
\label{20}
\fl \ma{V}\left(\left|\,\, x^i_{a\,,\,E}\,-\,\frac{1}{\hbar}\,\Theta^{ij}_E\,p^j_{a\,,\,E}\,-\, x^i_{b\,,\,B}\,\,\right|\right)\,\Psi(x^i_{a\,,\,E}\,,\,x^i_{b\,,\,B})\nonumber\\
\lo=\,\frac{1}{(2\pi)^{3/2}}\,\inte{k}\,\tilde{\ma{V}}(\vk)\,\rme^{\rmi\,k^i(x^i_{a\,,\,E}\,-
\,\frac{1}{\hbar}\,\Theta^{ij}_E\,p^j_{a\,,\,E}\,-\, x^i_{b\,,\,B})}\,\Psi(x^i_{a\,,\,E}\,,\,x^i_{b\,,\,B})\nonumber\\
\lo=\,\frac{1}{(2\pi)^{3/2}}\,\inte{k}\,\tilde{\ma{V}}(\vk)\,\rme^{\rmi\,\vk\cdot(\vx_{a\,,\,E}\,-
\,\vx_{b\,,\,B})}\,\rme^{-k^i\,\Theta^{ij}_E\,\partial^j_{\vx_{a,E}}}\,\Psi(x^i_{a\,,\,E}\,,\,x^i_{b\,,\,B})\nonumber\\
\lo=\,\ma{V}\left(\left|\,x^i_{a\,,\,E}\,-\,x^i_{b\,,\,B}\,\right|\right)
\,\rme^{\rmi\,\overleftarrow{\partial^i}_{\vx_{a,E}}\,
\Theta^{ij}_E\,\overrightarrow{\partial^j}_{\vx_{a,E}}}\,\Psi(x^i_{a\,,\,E}\,,\,x^i_{b\,,\,B})\nonumber\\
\lo=\,
\ma{V}\left(\left|\,x^i_{a\,,\,E}\,-\,x^i_{b\,,\,B}\,\right|\right)\,\star_{a,E}\,\Psi(x^i_{a\,,\,E}\,,\,x^i_{b\,,\,B})\,.
\eea

\noindent
where $\tilde{\ma{V}}(\vk)$ is the Fourier transform of $\ma{V}(|\vx|)$ and

\bea
\label{21}
\fl \ma{V}\left(\left|\,\vx_{a\,,\,E}\,-\,\vx_{b\,,\,B}\,\right|\right)\,\star_{a,E}\,\Psi(x^i_{a\,,\,E}\,,\,x^i_{b\,,\,B})\nonumber\\
\lo\equiv\,\ma{V}\left(\left|\,\vx_{a\,,\,E}\,-\,\vx_{b\,,\,B}\,\right|\right)\,\rme^{\rmi\,\overleftarrow{\partial^i}_{\vx_{a,E}}\,\Theta^{ij}\,\overrightarrow{\partial^j}_{\vx_{a,E}}}\,\Psi(x^i_{a\,,\,E}\,,\,x^i_{b\,,\,B})
\eea

\noindent
is the Gr\"onewold-Moyal or $\star$-product \cite{Gronewold,Moyal}. Notice that
$x^i_{a\,,\,E}$ does not commutes with $p^i_{a\,,\,E}$ but, however, it does commute
with $\Theta^{ij}_E \, p^j_{a\,,\,E}$ due to the antisymmetric character of
$\Theta^{ij}_E$. This observation is of the outmost importance for arriving at
Eq.(\ref{20}). Furthermore, since only the electron coordinates are sensitive to the
$\star$-product we drop, from now on, the subscript $E$ in this particular symbol. We single out

\bml
\label{22}
\bea
\int\,\rmd^3x\,\phi_{1}({\bf x})\star\phi_{2}({\bf x}) & = & \inte{x}\,\phi_{1}({\bf x})\phi_{2}({\bf x})\,,\label{mlett:a22}\\
\inte{x}\,\phi_{1}({\bf x})\star\phi_{2}({\bf x})\star\phi_{3}({\bf x}) & = & \inte{x}\,\phi_{3}({\bf x})\star\phi_{1}({\bf x})\star\phi_{2}({\bf x})\nonumber \\
 & = & \inte{x}\,\phi_{2}({\bf x})\star\phi_{3}({\bf x})\star\phi_{1}({\bf x})\,,\label{mlett:b22}
\eea
\eml

\noindent
as the properties of the $\star$-product \cite{Douglas1,Szabo1,Gomes1,Girotti1}
which will play a relevant role in our future developments.

We now address to the problem of computing ${\cal H}_{EB}$. By substituting
Eq.(\ref{20}) into Eq.(\ref{18}) one obtains,

\bea
\label{23}
\fl{\cal H}_{EB}\,\Psi(x^i_{c\,,\,E}\,,\,x^i_{d\,,\,B})\,&=&\,-\sum_{a=1}^{N}\int \rmd^3x_{b\,,\,B}\,n({\vx}_{b\,,\,B})\,\ma{V}\left(\left|\, x^i_{a\,,\,E}\,-\,\frac{1}{\hbar}\,\Theta^{ij}_E\,p^j_{a\,,\,E}\,-\, x^i_{b\,,\,B}\,\right|\right)\nonumber\\
&\times&\,\Psi(x^i_{c\,,\,E}\,,\,x^i_{d\,,\,B})\,\nonumber\\
&=&\,-\frac{N}{V}\sum_{a=1}^{N}\int \rmd^3x_{b\,,\,B}\,\ma{V}\left(\left|\, x^i_{a\,,\,E}\,-\, x^i_{b\,,\,B}\,\right|\right)\,\star_a\,\Psi(x^i_{c\,,\,E}\,,\,x^i_{d\,,\,B})\,\nonumber\\
&=&\,-\frac{N}{V}\sum_{a=1}^{N}\left[\int \rmd^3z\,\ma{V}(|\vz|)\right]\,\star_a\,\Psi(x^i_{c\,,\,E}\,,\,x^i_{d\,,\,B})\nonumber\\
&=&\,H_{EB}\,\Psi(x^i_{c\,,\,E}\,,\,x^i_{d\,,\,B})\,,
\eea

\noindent
which in view of the arbitrariness of $\Psi(x^i_{c\,,\,E}\,,\,x^i_{d\,,\,B})$ amounts to

\be
\label{24}
{\cal H}_{EB}\,=\,H_{EB}\,,
\ee

\noindent
as an operator identity. Thus, the noncommutativity of the electron position
observables does not affect the Hamiltonian $H_{EB}$. This is a consequence of the
continuous structure assumed for the ion background.

It remains to study the modifications induced by the noncommutativity on $H_E$. One
may convince oneself that

\be
\label{25}
{\cal H}_{E} \,=\, H_0\,+\,{\cal V}_E\,,
\ee

\noindent
where $H_0$ is given in Eq.(\ref{mlett:a8}), while

\bea
\label{26}
\fl{\cal V}_E\,=\,\frac{1}{2}\,\sum_{\vk_1\lambda_1}\sum_{\vk_2\lambda_2}
\sum_{\vk_3\lambda_3}\sum_{\vk_4\lambda_4}c^\dagger_{\vk_1\lambda_1}\,
c^\dagger_{\vk_2\lambda_2} \Bra{
{\vk}_{1}\lambda_{1}\,{\vk}_{2}\lambda_{2}}\ma{V}\left(\left|\vec{Q}_{a\,,\,E}-\vec{Q}_{b\,,\,E}\right|\right)\Ket{{\vk}_{3}
\lambda_{3}\,{\vk}_{4}\lambda_{4}}\nonumber\\
\times c_{\vk_4\lambda_4}\,c_{\vk_3\lambda_3}\,.\quad
\eea

Through standard manipulations \cite{Fetter}, the right hand side of Eq.(\ref{26})
can be written

\bea
\label{27}
\fl\Bra{{\vk}_{1}\lambda_{1}\,{\vk}_{2}\lambda_{2}}\ma{V}\left(\left|\,\vec{Q}_{a\,,\,E}\,-\,\vec{Q}_{b\,,\,E}\,\right|\right)
\Ket{{\vk}_{3}\lambda_{3}\,{\vk}_{4}\lambda_{4}}\nonumber\\
\lo=\,V^{-2}\,\left[\eta_{\lambda_{1}}(a)^\dagger\otimes\eta_{\lambda_{2}}(b)^\dagger\right]
\left[\eta_{\lambda_{3}}(a)\otimes\eta_{\lambda_{4}}(b)\right]
\,\inte{x_{a\,,\,E}}\inte{x_{b\,,\,E}}
\,\rme^{-\rmi\,\vk_{1}\cdot\vx_{a\,,\,E}}\nonumber\\
\times\,\rme^{-\rmi\,\vk_{2}\cdot\vx_{b\,,\,E}}\,\left[\ma{V}(|\,\vx_{a\,,\,E}\,-\,\vx_{b\,,\,E}\,|)\,\star_a\,\star_b\,
\rme^{\rmi\,\vk_{3}\cdot\vx_{a\,,\,E}}\,\rme^{\rmi\,\vk_{4}\cdot\vx_{b\,,\,E}}\right]\,,
\eea

\noindent
where

\be
\label{28}
\Phi_{\vk\lambda}(\vx)\,\equiv\,\Bracket{\vx}{\vk\lambda}\,=\,V^{-\frac{1}{2}}\,\rme^{\rmi\,\vx\cdot\vk}\,\eta_\lambda\,,
\ee

\noindent
is the free electron wave function, with

\bea
\label{29}
\eta_{\uparrow}\,=\,\left[\begin{array}{c}
1\\0\end{array}\right] \,,\quad
\eta_{\downarrow}\,=\,\left[\begin{array}{c}
0\\1\end{array}\right]\,.
\eea

\noindent
Furthermore,

\be
\label{30}
k_i\,=\,\frac{2\pi n_i}{L}\,,\quad i=1,2,3\quad\mathrm{and}\quad n_i=\pm 1,\,\pm
2,\,\ldots
\ee

\noindent
is the periodically quantized momentum. The use of Eqs.(\ref{mlett:a22}) and
\eref{mlett:b22} enable us to find

\bea
\label{31}
\fl\Bra{
{\vk}_{1}\lambda_{1}\,{\vk}_{2}\lambda_{2}}\ma{V}\left(\left|\,\vec{Q}_{a\,,\,E}\,-\,\vec{Q}_{b\,,\,E}\,\right|\right)
\Ket{{\vk}_{3}\lambda_{3}\,{\vk}_{4}\lambda_{4}}\nonumber\\
\lo=\,V^{-2}\,\delta_{\lambda_1\lambda_3}\,\delta_{\lambda_2\lambda_4}
\,\inte{x_{a\,,\,E}}\inte{x_{b\,,\,E}}\,\ma{V}(|\,\vx_{a\,,\,E}\,-\,\vx_{b\,,\,E}\,|)\nonumber\\
\times\,\left[\rme^{\rmi\,\vk_{3}\cdot\vx_{a\,,\,E}}\,\star_a\,\rme^{-\rmi\,\vk_{1}\cdot\vx_{a\,,\,E}}\right]\,
\left[\rme^{\rmi\,\vk_{4}\cdot\vx_{b\,,\,E}}\,\star_b\,\rme^{-\rmi\,\vk_{2}\cdot\vx_{b\,,\,E}}\right]\nonumber\\
\lo=\,V^{-1}\,\delta_{\vk_1+\vk_2\,,\,\vk_3+\vk_4}\,\delta_{\lambda_1\lambda_3}\delta_{\lambda_2\lambda_4}\,
\rme^{\rmi\,(\vk_3\wedge\vk_1\,+\,\vk_4\wedge\vk_2)}\,\inte{z}\,\ma{V}(|\vz|)\,\rme^{\rmi\,\vz\cdot(\vk_4\,-\,\vk_2)}\,,
\eea

\noindent
where the wedge product stands for

\be
\label{32} \vk\wedge\vp\,\equiv\, k^i\,\Theta^{ij}_E\, p^j\,.
\ee

\noindent
For arriving at the last term in the right hand side of Eq.(\ref{31}) we took advantage of

\be
\label{33}
\rme^{\rmi\,\vx\cdot\vk}\,\star\, \rme^{-\rmi\,\vx\cdot\vp}\,=\,\rme^{\rmi\,\vk\wedge\vp}\,
\rme^{\rmi\,\vx\cdot\vk}\,\rme^{-\rmi\,\vx\cdot\vp}\,.
\ee

The replacement of Eq.(\ref{31}) into Eq.(\ref{26}) yields

\bea
\label{34}
\fl{\cal V}_E\,=\,\frac{1}{2}\,\sum_{\vk_1\lambda_1}\sum_{\vk_2\lambda_2}
\sum_{\vk_3\lambda_3}\sum_{\vk_4\lambda_4}\,c^\dagger_{\vk_1\lambda_1}\,
c^\dagger_{\vk_2\lambda_2}\,V^{-1}\,\delta_{\vk_1+\vk_2\,,\,\vk_3+\vk_4}\,
\delta_{\lambda_1\lambda_3}\,\delta_{\lambda_2\lambda_4}\,
\rme^{\rmi\,(\vk_3\wedge\vk_1\,+\,\vk_4\wedge\vk_2)}\nonumber\\
\times\,\inte{z}\,\ma{V}(|\vz|)\,\rme^{\rmi\,\vz\cdot(\vk_4\,-\,\vk_2)}
\,c_{\vk_4\lambda_4} \,c_{\vk_3\lambda_3}\nonumber\\
\lo=\,\frac{1}{2V}\,\sum_{\vk\vp\vq}\sum_{\lambda_1\lambda_2}\,\left[\inte{z}\,\ma{V}(|\vz|)\,\rme^{\rmi\,\vq\cdot\vz}\right]
\,\rme^{-\rmi\,\vq\,\wedge\,(\vk\,-\,\vp)}\,c^\dagger_{\vk+\vq,\lambda_1}\,c^\dagger_{\vp-\vq,\lambda_2}\,
c_{\vp\lambda_2}\,c_{\vk\lambda_1}\,.
\eea

\noindent
This is the desired form of ${\cal V}_E$ in terms of creation and annihilation
operators. It  exhibits explicitly the noncommutativity. As is common practice, we
have chosen ${\vq}$ to designate the momentum transfer of the reaction
${\vp}\,+\,{\vk}\,\to\,({\vp}-{\vq})\,+\,({\vk}+{\vq})$.

As in the commutative case \cite{Fetter}, the contribution of the $\vq = 0$ mode in
the right hand side of Eq.(\ref{34}) cancels out those arising from ${\cal H}_{B}$
and ${\cal H}_{EB}$. This means that the noncommutativity does not destroy the
electric neutrality. Hence, the whole modified system collapses into

\be
\label{35}
{\cal H}\,=\,H_0\,+\,{\cal H}_I\,,
\ee

\noindent
where $H_0$ is given by Eq.(\ref{mlett:a8}), while ${\cal H}_I$ reads

\be
\label{36}
{\cal H}_I\,=\, \frac{2\pi\,e^2}{V}\,\sum_{\vec{k}\vec{p}\vec{q}}{}^{\prime}
\sum_{\lambda_1\lambda_2}\frac{\rme^{-\rmi\,\vq\,\wedge\,(\vk\,-\,\vp)}}{q^2}\,
c^{\dagger}_{\vec{k}+\vec{q}\,,\,\lambda_1}\,
c^{\dagger}_{\vec{p}-\vec{q}\,,\,\lambda_2}\,c_{\vec{p}\,,\,\lambda_2}\,
c_{\vec{k}\,,\,\lambda_1}\,.
\ee

\noindent
At this point a digression is in order. Notice that, in contradistinction with the
relativistic case, the commutative limit ($\Theta^{ij}_E \to 0$) in Eq.(\ref{36})
exists and is well defined.  To put it differently, the UV/IR mechanism
\cite{Seiberg}, that contaminates noncommutative relativistic field theories,
does not presently arise. This is of course due to the absence of ultraviolet
divergences in the nonrelativistic case. It is a rather simple exercise to verify
that ${\cal H}_I$ is Hermitean, as it must.

We have so far developed the tools to compute some of the physical effects induced
by the noncommutativity in the electron gas. We focus on the ground state energy
eigenvalue and employ, as in the commutative situation, time-independent
perturbation theory. We start by writing

\be
\label{37}
{\cal E}_0\,=\,E_{0}^{(0)}\,+\,\Bra{E^{(0)}_{0}}\,{\cal H}_I\,\Ket{E^{(0)}_{0}}\,+\,\sum_{i\ne0}\,\frac{\left|\Bra{E^{(0)}_{0}}\,{\cal {H}}_{I}\,\Ket{E^{(0)}_{i}}\right|^{2}}{E_{0}^{(0)}-E_{i}^{(0)}}\,+\,\cdots\,,
\ee

\noindent
where $\{E_{i}^{(0)}\}$ are the excited states of $H_0$. Since $H_0$ does not feel
the presence of noncommutativity, its eigenstates and corresponding eigenvalues
remain unchanged. Therefore, Eq.(\ref{9}) still holds true.

What come next is the computation of ${\cal E}^{(1)}_0$ which, according to
Eq.(\ref{37}), reads

\bea
\label{38}
\fl{\cal E}_0^{(1)}\, = \, \Bra{E^{(0)}_{0}}\,{\cal H}_I\,\Ket{E^{(0)}_{0}}\nonumber\\
\lo=\, \frac{2 \pi
\,e^{2}}{V}\sum_{\vk\vp\vq}{}^\prime\sum_{\lambda_{1}\lambda_{2}}\,\frac{
\rme^{-\rmi\,{\vq\,\wedge\,}({\vk}\,-\,{\vp})}}{q^{2}}\Bra{E^{(0)}_{0}}\,c^{\dagger}_{\vec{k}+\vec{q}\,,\,\lambda_1}\,
c^{\dagger}_{\vec{p}-\vec{q}\,,\,\lambda_2}\,c_{\vec{p}\,,\,\lambda_2}\,
c_{\vec{k}\,,\,\lambda_1}\,\Ket{E^{(0)}_{0}}\,.
\eea

\noindent
As already indicated, the mode $\vq = 0$ does not contribute to the right hand side
of Eq.(\ref{38}). Then, straightforward manipulations lead us to

\be
\label{39}
\fl\Bra{E^{(0)}_{0}}\,c^{\dagger}_{\vec{k}+\vec{q}\,,\,\lambda_1}\,
c^{\dagger}_{\vec{p}-\vec{q}\,,\,\lambda_2}\,c_{\vec{p}\,,\,\lambda_2}\,
c_{\vec{k}\,,\,\lambda_1}\,\Ket{E^{(0)}_{0}}\,=\,-\xi(k_{F}-p)\,\xi(k_{F}-k)\,\delta_{{\vp}-{
\vq},{\vk}}\,\delta_{\lambda_{1}\lambda_{2}}\,.
\ee

\noindent
Observe now that, when replacing Eq.(\ref{39}) into Eq.(\ref{38}), the factor
$\delta_{{\vp}-{ \vq},{\vk}}$ kills all noncommutative effects and, therefore,

\be
\label{40}
{\cal E}_0^{(1)}\,=\,E_0^{(1)}\,.
\ee

The computation of ${\cal E}^{(2)}_0$,

\be
\label{41}
{\cal E}^{(2)}_0\,=\,\sum_{i\ne0}\,\frac{\left|\Bra{E^{(0)}_{0}}\,{\cal {H}}_{I}\,\Ket{E^{(0)}_{i}}\right|^{2}}{E_{0}^{(0)}-E_{i}^{(0)}}\,,
\ee

\noindent
is cumbersome. We shall not pause here to present the details but merely mention
that it turns out to be given by

\be
\label{42}
{\cal E}^{(2)}_0\,=\,\frac{e^2}{2
a_0}\,N\,\left[\ep_0^{(2)r}\,+\,\ep_0^{(2)b}\left(\Theta\right)\right]\,.
\ee

\noindent
It is instructive to compare this result with its commutative counterpart, quoted in
Eqs.(\ref{11}) and (\ref{12}). On the one hand, $\ep_0^{(2)r}$ remains unaffected by
the noncommutativity while, on the other hand, the exchange correlation energy term,
$\ep_0^{(2)b}$, is modified as follows

\bea
\label{43}
\fl\ep_0^{(2)b}\to\ep_0^{(2)b}\left(\Theta\right)\nonumber\\
\fl=\,\frac{3}{16\pi^{5}}\int\frac{\rmd^{3}{\bf q}}{{\bf q}^{2}}\int_{|\vec{\bf
k}+\vec{\bf q}|>1}\rmd^{3}{\bf k}\int_{|\vec{\bf p}-\vec{\bf q}|>1}\rmd^{3}{\bf
p}\,\frac{\xi(1-{\bf k})\,\xi(1-{\bf p})\,\rme^{-2\,\rmi\,k_{F}^{2}\,\vec{\bf
q}\,\wedge\,(\vec{\bf k}\,-\,\vec{\bf p})}}{[\vec{\bf q}\cdot(\vec{\bf q}+\vec{\bf
k}-\vec{\bf p})](\vec{\bf q}+\vec{\bf k}-\vec{\bf p})^{2}}\,.
\eea

\noindent
One may easily verify that $\ep_0^{(2)b}\left(\Theta\right)$ is real, as demanded by
the hermiticity of ${\cal H}_I$. This allows the replacement of the exponential by
its real part, i.e.,

\bea
\label{44}
\fl\ep_0^{(2)b}\left(\Theta\right)\nonumber\\
\fl=\,\frac{3}{16\pi^{5}}\int\frac{\rmd^{3}{\bf q}}{{\bf q}^{2}}\int_{|\vec{\bf
k}+\vec{\bf q}|>1}\rmd^{3}{\bf k}\int_{|\vec{\bf p}-\vec{\bf q}|>1}\rmd^{3}{\bf
p}\,\frac{\xi(1-{\bf k})\,\xi(1-{\bf p})\,\cos\left[2\,k_{F}^{2}\,\vec{\bf
q}\wedge(\vec{\bf k}-\vec{\bf p})\right]}{[\vec{\bf q}\cdot(\vec{\bf q}+\vec{\bf
k}-\vec{\bf p})](\vec{\bf q}+\vec{\bf k}-\vec{\bf p})^{2}}\,.\quad
\eea

\noindent
Needless to say, the equivalence between Eqs.(\ref{43}) and (\ref{44}) can also be
checked by direct computation. Since the argument of the trigonometric function
depends on $k_{F}$ and, therefore, on $V$, the exchange correlation energy is no
longer a constant. As consequence, the thermodynamics properties are modified by the
noncommutativity.

We have not been able of computing analytically the integral in Eq.(\ref{44}). To
proceed further on, we assume that the global features of the system are insensitive
to the direction of the vector $\theta^i \equiv 1/2 \ep^{ijk} \Theta^{jk}_E$. We
may, then, replace the right hand side of Eq.(\ref{44}) by its average over all
possible directions of ${\vec \theta}$. This is effectively achieved by integrating
over the angles of ${\vec \theta}$ which yields

\bea
\label{45}
\fl\epsilon_{0}^{(2)b}(\theta)\, \equiv \, \frac{1}{4\pi}\int \rmd\Omega_{{\vec\theta}}\,\epsilon_{0}^{(2)b}({\vec \theta})\nonumber \\
\lo=\,\frac{3}{16\pi^{5}}\int\frac{\rmd^{3}{\bf q}}{{\bf q}^{2}}\int_{|\vec{\bf k}+\vec{\bf q}|>1}\rmd^{3}{\bf k}\int_{|\vec{\bf p}-\vec{\bf q}|>1}\rmd^{3}{\bf p}\,\frac{\xi(1-{\bf k})\,\xi(1-{\bf p})}{[\vec{\bf q}\cdot(\vec{\bf q}+\vec{\bf k}-\vec{\bf p})](\vec{\bf q}+\vec{\bf k}-\vec{\bf p})^{2}}\nonumber \\
\times\,\frac{\sin(k_{F}^{2}\theta|\vec{\bf q}\times(\vec{\bf k}-\vec{\bf
p})|)}{k_{F}^{2}\theta|\vec{\bf q}\times(\vec{\bf k}-\vec{\bf p})|}\,.
\eea

\noindent
Let us concentrate on analyzing different limiting cases. For $\theta = |{\vec
\theta}| = 0$ one returns unambiguously to the commutative model. On the other hand,
when $\theta \to \infty \Longrightarrow \epsilon_{0}^{(2)b} \to 0$ implying that
${\cal E}_0^{(2)} < E_0^{(2)}$. However, thermodynamic quantities, such as pressure
and bulk modulus, will remain unaltered because the difference ${\cal E}_0^{(2)} -
E_0^{(2)}$ is just a constant.

The next step consists in bringing into play the Seiberg-Witten \cite{Seiberg1} map.
By expanding the trigonometric function in Eq.(\ref{45}) around $\theta = 0$ one
arrives at

\be
\label{46}
\epsilon_{0}^{(2)b}(\theta)\,=\,\epsilon_{0}^{(2)b} \,-\,\frac{1}{32 \pi^5}\,k_F^4\,R\,\theta^2\,+\,{\cal O}\left( \theta^4\right)\,,
\ee

\noindent
where

\be
\label{47}
\fl R\,=\,\int\frac{\rmd^{3}{\bf q}}{{\bf q}^{2}}\int_{|\vec{\bf k}+\vec{\bf
q}|>1}\rmd^{3}{\bf k}\int_{|\vec{\bf p}-\vec{\bf q}|>1}\rmd^{3}{\bf p}\,\frac{\xi(1-{\bf
k})\,\xi(1-{\bf p})\,|\vec{\bf q}\times(\vec{\bf k}-\vec{\bf p})|^2}{[\vec{\bf
q}\cdot(\vec{\bf q}+\vec{\bf k}-\vec{\bf p})](\vec{\bf q}+\vec{\bf k}-\vec{\bf
p})^{2}}\,.
\ee

\noindent
The convergence of the ${\bf k}$ and ${\bf p}$ integrals is secured by the fact that
they run over finite intervals. On the other hand, power counting tell us that the
improper ${\bf q}$ integral also converges. Hence, $R$ exists and is well defined.
The situation changes drastically for those integrals that act as coefficients of
higher orders in $\theta$. There, power counting indicates that they are divergent.
The way out from the trouble consists in carrying out the ${\bf q}$ integral between
$0$ and $\Lambda$ being $\Lambda$ a cutoff such that, as $\theta \to 0$, $1/k_F^2
\theta$ goes to infinity faster than $\Lambda$.

By collecting all the results, Eq.(\ref{37}) yields

\bea
\label{48}
\fl{\cal E}_0\,=\, \frac{e^{2}}{2a_{0}}N\left[\frac{2.21}{r_{s}^{2}}-\frac{0.916}{r_{s}}+0.0622\ln r_{s}-0.094-\frac{1}{32\pi^{5}}k_{F}^{4}\theta^{2}R\,+\,{\cal O}(\theta^{4},r_{s}\ln r_{s})\right]\nonumber\\
\lo=E_0\,-\,N\,\frac{m}{\hbar}\frac{3}{32\pi^{5}}\,k_{F}^{4}\,e^2\,R\,\theta^{2}\,+\,{\cal
O}(\theta^{4})\,.
\eea

\noindent
The noncommutativity certainly modifies the ground state energy and, as consequence,
the ionization potential of the material being treated as an electron gas. Moreover,
from the comparison of Eq.(\ref{48}) with the commutative electron gas at non zero
temperature \cite{Gellmann2} one may conclude that $\theta$ acts as a reference
temperature \cite{Girotti2} since $\partial {\cal E}_0/\partial \theta$ is a linear
function of $\theta$ much as the specific heat at constant volume is a linear
function of the absolute temperature. This is the main outcome of the present paper, namely, the noncommutativity of the position observables lifts the degeneracy of the model and can be interpreted as if the electron gas would be at nonzero temperature.

\ack
Both authors thank Dr. Alysson F. Ferrari for useful discussions.

This work was partially supported by Conselho Nacional de Desenvolvimento Cient\'\i
fico e Tecnol\'ogico (CNPq) and by PRONEX under contract CNPq 66.2002/1998-99.

\end{document}